\documentclass[12pt]{article}
\usepackage[utf8]{inputenc}
\usepackage[russian,english]{babel} 
\usepackage[T2A]{fontenc}   
\usepackage{mathtools}
\usepackage{amsfonts,amssymb,amsmath} 
\usepackage{graphicx} 
\usepackage{cite}
\usepackage{hyperref}
\hypersetup{colorlinks=true, linkcolor=blue, filecolor=blue, urlcolor=blue} 

\usepackage{epstopdf} 

\usepackage[table]{xcolor} 

\usepackage{diagbox} 

\newcommand{\der}[2]{\dfrac{\mathrm{d} #1}{\mathrm{d} #2}}
\def\L{\mathcal{L}}     
\def\D{\mathcal{D}}     

\begin{document}

\title{Generation of C-NOT, SWAP, and C-Z Gates \\
for Two Qubits Using Coherent 
and Incoherent \\Controls and Stochastic Optimization}

\author{Oleg~V.~Morzhin$^{1,}$\footnote{E-mail: \url{morzhin.oleg@yandex.ru};~ 
    \href{http://www.mathnet.ru/eng/person30382}{mathnet.ru/eng/person30382};~ 
    \href{https://orcid.org/0000-0002-9890-1303}{ORCID 0000-0002-9890-1303}} 
    \quad and \quad 
Alexander~N.~Pechen$^{1,}$\footnote{Corresponding author. E-mail: \url{apechen@gmail.com};~ 
    \href{http://www.mathnet.ru/eng/person17991}{mathnet.ru/eng/person17991};~ 
    \href{https://orcid.org/0000-0001-8290-8300}{ORCID 0000-0001-8290-8300}} 
    \vspace{0.2cm} \\
$^1$ Department of Mathematical Methods for Quantum Technologies, \\
Steklov Mathematical Institute of Russian Academy of Sciences, \\
8~Gubkina Str., Moscow, 119991, Russia \\
(http://www.mi-ras.ru/eng/dep51)}

\date{}
\maketitle

\begin{abstract}
In this work, we consider a general form of the dynamics of open quantum systems 
determined by the Gorini--Kossakowsky--Sudarchhan--Lindblad type 
master equation with simultaneous coherent and incoherent controls with three particular forms of the 
two-qubit Hamiltonians. Coherent control enters   
in the Hamiltonian and incoherent control enters 
in both the Hamiltonian 
and the superoperator of dissipation. For these systems, 
we analyze the control problems of generating 
two-qubit C-NOT, SWAP, and C-Z gates using 
with piecewise constant controls and stochastic optimization 
in the form of an adapted version of the dual annealing algorithm. In the numerical experiment, we analyze the minimal infidelity obtained by the dual annealing for various values of 
strength of the interaction between the system and the environment.

\vspace{0.3cm}

\end{abstract} 

\section{Introduction}  

Optimal control of quantum systems such as atoms or molecules is an important
research direction at the intersection 
of mathematics, physics, chemistry, and technology. It aims to obtain 
desired properties of the controlled quantum system using 
external variable influence.  Quantum control 
is crucial for development of various quantum technologies such as
quantum computing, NMR, laser chemistry, etc.~\cite{DongPetersenBook2023, KuprovBook2023, 
KochEPJQuantumTechnol2022, AlessandroBook2021, 
KurizkiKofmanBook2021, KwonTomonagaEtAl2021, 
BaiChenWuAn2021, Acin2018, KochJPhysCondensMatter2016, 
DongWuYuanLiTarn2015, CongBook2014, AltafiniTicozzi2012, 
BonnardBook2012, Gough_2012, Shapiro_Brumer_Book_2ndEd_2012, 
BrifNewJPhys2010, FradkovBook2007, LetokhovBook2007, 
TannorBook2007, Rice_Zhao_2000, ButkovskyBook1990}. 

In general, in quantum control various types of optimization tools 
are used including Pontryagin maximum principle~(\cite[Ch.~4]{ButkovskyBook1990}, \cite{BoscainPRXQuantum2021}), 
Krotov type methods \cite{Kazakov_Krotov_1987, Tannor_Kazakov_Orlov_1992, 
Bartana_Kosloff_Tannor_2001, Jager_Reich_et_al_2014, Goerz_NJP_2014_2021, 
Morzhin_Pechen_UMN_2019,MorzhinPechenBulletinIrkutsk2023}), Hamilton--Jacobi--Bellmann equation~\cite{Gough_Belavkin_Smolyanov_2005}, 
Zhu--Rabitz \cite{Zhu_Rabitz_1998} method, 
GRadient Ascent Pulse Engineering (GRAPE) 
\cite{KhanejaGRAPE2005, Jager_Reich_et_al_2014, PetruhanovPhotonics2023_H_T_gates, PetruhanovPechenJPA2023} 
operating with piecewise constant controls, gradients, matrix exponentials, 
conditional gradient method~\cite{Morzhin_Pechen_LJM_2019},  
gradient projection 
methods~\cite{Morzhin_Pechen_LJM_2019, MorzhinPechenBulletinIrkutsk2023, MorzhinPechenQIP2023}, 
speed-gradient method~\cite{Pechen_Borisenok_Fradkov_2022}, 
gradient-free Chopped RAndom Basis (CRAB) ansatz \cite{CanevaPRA2011, DoriaCalarcoMontangero2011, MullerSaidJelezkoCalarcoMontangero2022} with coherent control defined using trigonometric functions, 
genetic algorithms~\cite{JudsonRabitz1992, PechenRabitzPRA2006, BrownPaternostroFerraro2023}, 
dual annealing~\cite{MorzhinPechenBulletinIrkutsk2023}, machine learning~\cite{Dong_Chen_Tarn_Pechen_Rabitz_2008, 
Niu_Boixo_Smelyanskiy_et_al_2019}, etc.

From the experimental point of view, quantum systems are often open, 
i.e. they interact with their environment. On one hand, this 
is considered as an~obstacle for controlling quantum systems. 
However, in some cases one can try to use the environment 
as a useful control resource --- in this work we exploit the {\it incoherent control} approach which was proposed in~\cite{PechenRabitzPRA2006, PechenPRA2011},  
where spectral, generally time-dependent and non-equilibrium, density of incoherent photons was used as control 
jointly with {\it coherent control} by lasers. 
Following this approach, various types and aspects 
of optimal control problems for one- and two-qubit systems 
were analyzed recently in our works~\cite{PetruhanovPhotonics2023_H_T_gates, Pechen_Petruhanov_Morzhin_Volkov_CNOT_CPHASE_gates, Morzhin_Pechen_LJM_2019, MorzhinPechenQIP2023},~etc. In this approach, the dynamics (both for Markovian, including in the Born-Markov approximation, and non-Markovian cases) of an $N$-level quantum system is considered with general master equation under simultaneous coherent~$u$ and 
incoherent~$n$ controls:
\begin{equation}
\label{General_Master_Equation_with_controls}
\der{\rho(t)}{t} = \L^{u, n}_t (\rho(t)) := -i [H^{u, n}_t, \rho(t)] + \varepsilon\underbrace{\sum_k\gamma_k(t) \D_k (\rho(t))}_{\displaystyle \D^n_t (\rho(t))},\quad \rho(0) = \rho_0. 
\end{equation}
Here $\rho(t) : \mathcal{H} \to \mathcal{H}$ is $N \times N$ density matrix (positive semi-definite, $\rho(t) \geq 0$, with unit trace, ${\rm Tr}\rho(t) = 1$); $H_t^{u,t}$ is the Hamiltonian which represents either memoryless Markovian or having memory terms non-Markovian case; the coefficient $\varepsilon > 0$ determines strength of the interaction between the system and the environment; the last term in is the dissipator (superoperator of dissipation) depending on incoherent control~$n$ where $\{\gamma_k\}$ denote decoherence rates depending on time; $\rho_0$ is a~given initial density matrix. In this work we consider the Planck's constant $\hbar=1$. 

Basic objects in quantum information and computing are
qubits, qutrits, qudits and quantum gates (e.g., the recent 
papers~\cite{Roy_Li_Kapit_Schuster_2023, 
KwonTomonagaEtAl2021, Wang_Hu_Sanders_Kais_2020}, 
also~\cite{Kurgalin_Borzunov_Book_2021, Holevo_Book_2019, NielsenChuangBook2010, Valiev_Uspekhi_2005, Palao_Kosloff_2003}, etc.).
Mathematical modeling of physical processes 
for generating quantum gates involves, e.g., 
the Schr\"{o}dinger equation with coherent control 
in the Hamiltonian, 
the Gorini--Kossakowski--Sudarshan--Lindblad (GKSL) master 
equation with coherent (in the Hamiltonian) and incoherent 
(in the Hamiltonian via the Lamb shift and 
in the superoperator of dissipation) 
controls~\cite{Goerz_NJP_2014_2021, PetruhanovPhotonics2023_H_T_gates, 
Fernandes_Fanchini_deLima_Castelano_2023} and 
various objective functionals for 
characterizing how well certain unitary quantum gate is generated.  

Proposed in~\cite{Goerz_NJP_2014_2021} approach 
for modelling generation of ~$M$-qubit quantum gates ($M \geq 1$) considers dissipative dynamics depending only on coherent control and uses only three initial density matrices instead of the corresponding full basis (e.g., three matrices instead of sixteen for the two-qubit case). 
While the recent work~\cite{Fernandes_Fanchini_deLima_Castelano_2023} 
considers only coherent control (similar to~\cite{Goerz_NJP_2014_2021}), the recent works~\cite{PetruhanovPhotonics2023_H_T_gates, Pechen_Petruhanov_Morzhin_Volkov_CNOT_CPHASE_gates} extend the approach of~\cite{Goerz_NJP_2014_2021} to the GKSL-type quantum systems with simultaneous coherent control and incoherent control (in the approach of~\cite{PechenRabitzPRA2006, PechenPRA2011}) 
for generating various one- and two-qubit quantum gates
(H, T, C-NOT, C-PHASE including C-Z). As in~\cite{Pechen_Petruhanov_Morzhin_Volkov_CNOT_CPHASE_gates}, below we use the abbreviation ``GRK'' meaning the Goerz--Reich--Koch approach. 

In this work, on one hand similarly  to~\cite{Pechen_Petruhanov_Morzhin_Volkov_CNOT_CPHASE_gates} we consider the same three versions of the two-qubit quantum system, the same approach with piecewise constant controls, one of the objective functionals and DAA. In difference, we consider SWAP gate in addition to C-NOT and C-Z gates, and obtain the DAA results for C-NOT, SWAP, C-Z under different DAA settings and upper constraint for incoherent controls for analyzing dependence of the objective values on the parameter $\varepsilon$ determining strength of the interaction between the system and the environment. The DAA implementation~\cite{scipy_dual_annealing} is adapted.

The structure of the paper is the following. Sec.~\ref{section2} contains the formulation of the considered three types of open quantum systems and one type of the GRK objectives which is used here for C-NOT, SWAP, C-Z gates. Sec.~\ref{section3} describes the DAA optimization approach and the corresponding numerical results with respect to varying~$\varepsilon$. Conclusions section~\ref{Section_Conclusion} resumes the article. 

\section{Two-Qubit Quantum Systems and Objective Functionals}
\label{section2}

\subsection{Quantum Systems with the Three Kinds of the Hamiltonian}

For two qubits ($M=2$), the Hilbert space is
$\mathcal{H} =  \mathbb{C}^2 \otimes \mathbb{C}^2\approx\mathbb C^4$, i.e. $N=4$. As in~\cite{Pechen_Petruhanov_Morzhin_Volkov_CNOT_CPHASE_gates, MorzhinPechenBulletinIrkutsk2023, MorzhinPechenQIP2023}, consider the corresponding Markovian two-qubit case of~(\ref{General_Master_Equation_with_controls}).
The Hamiltonian $H^{u, n}_t$ for each of the three systems ($k=1,2,3$) has the following general form: 
\begin{align}
H^{u,n}_{t,k}= H_{S,k} + \varepsilon H^n_{{\rm eff}, t} + V_k u(t).
\label{general_form_of_Hamiltionian}
\end{align}
Here scalar coherent control $u$ and vector incoherent control $n=(n_1, n_2)$ are considered, in general, as piecewise continuous on $[0, T]$ and piecewise constant in the optimization DAA approach used in this article, control $f=(u,n)$;
$H_{S,k}$ is the free Hamiltonian; $H_{{\rm eff}, n(t)}$ is the effective Hamiltonian depending on values of incoherent control  which describes the Lamb shift;  $V_k$ is a~given Hermitian matrix which determines interaction with $u(t)$; the coefficient $\varepsilon > 0$. Below in this article, we change $\varepsilon$ for generating a~series of instances for each $H^{u,n}_{t,k}$, $k=1,2,3$.  

We denote by $\mathbb{I}_2$ the $2\times 2$ identity matrix, and denote Pauli matrices and uppering, lowering one-qubit matrices as
\[
\sigma_x = \begin{pmatrix}
0 & 1 \\
1 & 0
\end{pmatrix},\quad
\sigma_y = \begin{pmatrix}
0 & -i \\
i & 0
\end{pmatrix},\quad
\sigma_z = \begin{pmatrix}
1 & 0 \\
0 & -1
\end{pmatrix},\quad \sigma^+ = \begin{pmatrix}
0 & 0 \\ 1 & 0
\end{pmatrix}, \quad
\sigma^- = \begin{pmatrix}
0 & 1 \\ 0 & 0
\end{pmatrix}.
\]

The effective Hamiltonian in  (\ref{general_form_of_Hamiltionian}) and the  superoperator of dissipation are the same for all three considered systems:
\begin{align}
H^n_{{\rm eff},t} &=
\Lambda_1 n_1(t) \left( \sigma_z \otimes \mathbb{I}_2 \right) + 
\Lambda_2 n_2(t) \left( \mathbb{I}_2 \otimes \sigma_z \right), 
\label{effective_Hamiltonian} \\
\mathcal{D}^{n}(\rho) &= 
\mathcal{D}^{n_1}(\rho) + 
\mathcal{D}^{n_2}(\rho), 
\label{dissipator}
\\
\mathcal{D}^{n_j}_t(\rho) &=  \Omega_j (n_j(t) + 1) \left( 2 \sigma^-_j \rho \sigma^+_j - 
\sigma_j^+ \sigma_j^- \rho - \rho \sigma_j^ + \sigma_j^- \right) + \nonumber \\
&+ \Omega_j n_j(t) \left( 2\sigma^+_j \rho \sigma^-_j - 
\sigma_j^- \sigma_j^+ \rho - \rho \sigma_j^- \sigma_j^+ \right),
\qquad j = 1,2,\nonumber 
\end{align}
where $\Lambda_j>0$ and $\Omega_j>0$ are some constants depending on the details of interaction between the system and the environment, $\sigma_1^{\pm} = \sigma^{\pm} \otimes \mathbb{I}_2$, and $\sigma_2^{\pm} = \mathbb{I}_2 \otimes \sigma^{\pm}$.

For the general form~(\ref{general_form_of_Hamiltionian}), we consider the following three variants which produce the corresponding three quantum systems. 

{\bf System 1.} The free and interaction Hamiltonians are
\begin{align*} 
H_{S,1} &=  \frac{\omega_1}{2} \left( \sigma_z \otimes \mathbb{I}_2 \right) 
+ \frac{\omega_2}{2} \left( \mathbb{I}_2 \otimes \sigma_z \right),\\
V_1 &= \sigma_x \otimes \mathbb{I}_2 + \mathbb{I}_2 \otimes \sigma_x.  
\end{align*}

{\bf System 2.} The free and interaction Hamiltonians are
\begin{align*} 
H_{S,2} &= H_{S,1} =\frac{\omega_1}{2} \left( \sigma_z \otimes \mathbb{I}_2 \right) 
+ \frac{\omega_2}{2} \left( \mathbb{I}_2 \otimes \sigma_z \right),\\
V_2 &= \sigma_x \otimes \sigma_x. 
\end{align*}

{\bf System 3.} The free and interaction Hamiltonians are
\begin{align*}
H_{S,3} &=
\sigma_z \otimes {\mathbb I}_2 + {\mathbb I}_2 \otimes \sigma_z + 
\alpha (\sigma_y \otimes \sigma_y + \sigma_z \otimes \sigma_z), \quad \alpha>0,\\
V_3 &= \sigma_x \otimes {\mathbb I}_2. 
\end{align*}  

We consider the following constraints on controls:
\begin{align}
\label{constraints_on_controls}
f(t) = (u(t), n(t)) \in [-u_{\max}, u_{\max}] \times [0, n_{\max}]^2, \quad t \in [0, T],
\end{align}
where $u_{\max},~n_{\max}>0$.

As in~\cite{Pechen_Petruhanov_Morzhin_Volkov_CNOT_CPHASE_gates, MorzhinPechenBulletinIrkutsk2023, MorzhinPechenQIP2023}, consider the following realification (parameterization) of the two-qubit density matrix:
\begin{align*}
\rho = \begin{pmatrix}
\rho_{1,1} & \rho_{1,2} & \rho_{1,3} & \rho_{1,4} \\
\rho_{1,2}^{\ast} & \rho_{2,2} & \rho_{2,3} & \rho_{2,4} \\
\rho_{1,3}^{\ast} & \rho_{2,3}^{\ast} & \rho_{3,3} & \rho_{3,4} \\
\rho_{1,4}^{\ast} & \rho_{2,4}^{\ast} & \rho_{3,4}^{\ast} & \rho_{4,4} 
\end{pmatrix} 
= \begin{pmatrix}
x_1 & x_2 + i x_3 & x_4 + i x_5 & x_6 + i x_7 \\
x_2 - i x_3 & x_8 & x_9 + i x_{10} & x_{11} + i x_{12} \\
x_4 - i x_5 & x_9 - i x_{10} & x_{13} & x_{14} + i x_{15} \\
x_6 - i x_7 & x_{11} - i x_{12} & x_{14} - i x_{15} & x_{16}  
\end{pmatrix}.
\end{align*}
Here $x_j \in \mathbb{R}$, $j=\overline{1,16}$. The 
condition ${\rm Tr}\rho=1$ implies the linear condition $x_1+x_8+x_{13}+x_{16}=1$. Using the realification for the three two-qubit systems of~
(\ref{General_Master_Equation_with_controls}) with 
(\ref{general_form_of_Hamiltionian}),
(\ref{effective_Hamiltonian}), 
(\ref{dissipator}), inin~\cite{Pechen_Petruhanov_Morzhin_Volkov_CNOT_CPHASE_gates} we consdiered two-qubit gate generation using DAA and GRAPE. 

\subsection{Goerz--Reich--Koch Type Objective Functionals for the Problems of Generating C-NOT, SWAP, and C-Z Gates} 

For the two-qubit systems ($N=2$), consider
the problems of generation of the following
target unitary gates: C-NOT (controlled NOT), 
SWAP, and $\text{C-Z} = \text{C-PHASE}(\pi)$ (controlled $\pi$ phase) 
gates. These gates are defined in the computational basis by the unitary matrices
\begin{align*}
\textrm{C-NOT} = \begin{pmatrix}
1 & 0 & 0 & 0\\
0 & 1 & 0 & 0\\
0 & 0& 0 & 1\\
0 & 0 & 1 & 0
\end{pmatrix},
\quad 
\textrm{SWAP} = \begin{pmatrix}
1 & 0 & 0 & 0\\
0 & 0 & 1 & 0\\
0 & 1& 0 & 0\\
0 & 0 & 0 & 1
\end{pmatrix},
\quad
\textrm{C-Z} = \begin{pmatrix}
1 & 0 & 0 & 0\\
0 & 1 & 0 & 0\\
0 & 0& 1 & 0\\
0 & 0 & 0 & -1
\end{pmatrix}.
\end{align*} 

Following the GRK approach, for the problem of generating such a~two-qubit gate, we consider --- instead of the $(N^2 = 16)$-dimensional set of states --- the proposed in~\cite{Goerz_NJP_2014_2021} minimal set consisting just of the three states which, for the two-qubit case, have the form
\begin{align}
\label{three_initial_states}
\rho_{0,1} = {\rm diag}\left(\frac{2}{5}, \frac{3}{10}, \frac{1}{5}, \frac{1}{10} \right), \quad \rho_{0,2} = \frac{1}{4}J_4, \quad \rho_{0,3} = \frac{1}{4}\mathbb{I}_4,
\end{align}
where $J_4$ denotes the $4 \times 4$ matrix whose all elements are equal to~1. Further, the objective functional to be minimized is defined by the mean value of the three squared Hilbert--Schmidt distances multiplied by $1/2$, each of them is considered between the $m$th final density matrix, $\rho_m(T)$ (corresponds to the $m$th special initial density matrix, $\rho_{0,m}$), and the target density matrix $\rho_{{\rm target},m} = U \rho_{0,m} U^{\dagger}$: 
\begin{align}
F_U^{\rm GRK,sd}(f) = \frac{1}{6}\sum_{m = 1}^3 \|\ \rho_m(T) - U \rho_{0,m} U^{\dagger} \|^2 \to \inf.
\label{objective_functional}
\end{align}

The objective $F_U^{\rm GRK,sd}$ in principle is the same that the formula~(A1) in~\cite{Goerz_NJP_2014_2021}, but we introduce the multipliers $1/2$ and $1/3$. 
The set of initial states~(\ref{three_initial_states}) is a~particular case for the 
general formulas (4a), (4b), (4c) given in the same~article and also in the dissertation~\cite[p.~71]{Reich_Dissertaion_2015}. For brevity, we sometimes call the objective to be minimized in this work as infidelity.

Table~1 shows for each of the three considered gates that $U \rho_{0,m} U^{\dagger}$:
\begin{itemize}
\item does not change all the three initial states $\rho_{0,m}$, $m=1,2,3$;
\item changes only one initial state: if $U$ is C-NOT or SWAP, then the first initial state is changed, while if $U$ is C-Z, then the second initial state is changed.
\end{itemize}

\begin{table}[ht!]  
\centering
\caption{The matrices $U \rho_m(0) U^{\dagger}$ for $m=1,2,3$ when $U$ is either C-NOT, SWAP or C-Z. Color highlights cases when action of the gate on the matrix is non-trivial.}

\begin{tabular}{|c|c|c|c|}
\hline
\diagbox[width=10em]{Gate, $U$}{\hspace{.1cm} Target state} & $U \rho_{0,1} U^{\dagger}$ & $U \rho_{0,2} U^{\dagger}$ & $U \rho_{0,3} U^{\dagger}$ \\ 
\hline
C-NOT & \cellcolor{cyan!10}${\rm diag}\left(\frac{2}{5}, \frac{3}{10}, \frac{1}{10}, \frac{1}{5} \right) \neq \rho_{0,1}$ & $\frac{1}{4}J_4 = \rho_{0,2}$ & $\frac{1}{4}\mathbb{I}_4 = \rho_{0,3}$ \\ \hline
SWAP & \cellcolor{cyan!10}${\rm diag}\left(\frac{2}{5}, \frac{1}{5}, \frac{3}{10}, \frac{1}{10} \right) \neq \rho_{0,1}$
& $\frac{1}{4}J_4 = \rho_{0,2}$ & $\frac{1}{4}\mathbb{I}_4 = \rho_{0,3}$ \\ \hline
C-Z & $ {\rm diag}\left(\frac{2}{5}, \frac{3}{10}, \frac{1}{5}, \frac{1}{10} \right)= \rho_{0,1}$
& \cellcolor{cyan!10}$\frac{1}{4} \left(\begin{smallmatrix}
1 & 1 & 1 & -1 \\
1 & 1 & 1 & -1 \\
1 & 1 & 1 & -1 \\
-1 & -1 & -1 & 1 
\end{smallmatrix}\right) \neq \rho_{0,2}$  & $\frac{1}{4}\mathbb{I}_4 = \rho_{0,3}$ \\ \hline
\end{tabular}  
\end{table} 

\section{Numerical Approach and Results}
\label{section3}

\subsection{Approach Based on Piecewise Constant Controls and Dual Annealing} 

Here we define the class of admissible piecewise constant coherent 
and incoherent controls at the uniform grid introduced on 
$[0,T]$ with $K$ equal subintervals, i.e. 
\begin{align}
\label{piecewise_constant_controls}
u(t) = \sum\limits_{i=1}^{K} \theta_{[t_i, t_{i+1})}(t) u^i, 
\quad 
n_j(t) = \sum\limits_{i=1}^{K} \theta_{[t_i, t_{i+1})}(t) n^i_j, \quad j = 1,2,
\end{align}
and with fixed $u_{\max}$, $n_{\max}$ 
in~(\ref{constraints_on_controls}). In~(\ref{piecewise_constant_controls}), we consider $t_1 = 0$, $t_{i+1} - t_i = T/K$, $\theta_{[t_i, t_{i+1})}(t)$, and the characteristic function which returns~1, if 
$t \in [t_i, t_{i+1})$, and returns~0, if otherwise. For the final time~$t = t_{K+1} = T$, we continue the last values: $u(T) = u(T-)$, $n_j(T) = n_j(T-)$. The DAA search is performed in the parallelepiped $[-u_{\max}, u_{\max}]^{K} \times [0, n_{\max}]^{2K}$. 

Thus, instead of optimization work in the infinite-dimensional space, here we should perform finite-dimensional optimization work where the control parameters 
\[
\{u^1, \dots, u^K, ~ n_1^1, \dots, n_1^K, ~ n_2^1, \dots, n_2^K\}
\]
are considered in the $3K$-dimensional parallelepipedal domain.  

DAA is a~zeroth-order stochastic tool as genetic algorithms, differential evolution, particle-swarm optimization, sparrow search algorithm, etc. It tries to find a~global minimizer of an~objective function without using its gradient or/and Hessian. As~\cite{scipy_dual_annealing} informs, the corresponding implementation of DAA in {\tt SciPy} represents the combination of the simulated annealing from~\cite{Tsallis_1988, Tsallis_Stariolo_1996} and the local search strategy~\cite{Xiang_et_al_1997, Xiang_Gong_2000}.  

We use DAA with the default settings~\cite{scipy_dual_annealing} except of the following three items: 1)~the default {\tt initial\_temp} (initial artificial temperature) is increased to $3 \times 10^4$, i.e. near in six times; 2)~{\tt maxfun} (which determines the number of objective calls) is decreased from the default $10^7$ to $3 \times 10^4$; 3)~{\tt maxiter} is increased from the default $10^3$ to $3 \times 10^3$. With respect to the algorithmic parameter {\tt initial\_temp}, we read 
``higher values to facilitates a~wider search ... allowing dual\_annealing to escape local minima that it is trapped in'' in~\cite{scipy_dual_annealing}. 

Taking into account the stochastic nature of DAA, we perform for the same optimization problem several trials of DAA and then compare the computed value of the objective over the trials. For speed up the DAA work, it was performed in the parallel and sequential computations. For the numerical results, their storing and visualization are done, correspondingly, with help of the Python libraries {\tt sqlite3} and {\tt Matplotlib}.

In contrast to the use of DAA in~\cite{Pechen_Petruhanov_Morzhin_Volkov_CNOT_CPHASE_gates}, here we do the following changes: 
\begin{itemize}
\item we set $K = 200$ instead of $K=100$;
\item we set $n_{\max} = 20$ instead of $n_{\max} = 10$ and $u_{\max} = 20$ instead of $u_{\max} = 30$;
\item for each particular optimal control problem, we perform ten DAA trials instead of three DAA trials;
\item we vary $\varepsilon$ along the set~(\ref{values_of_epsilon}) instead of considering only $\varepsilon = 0.1$;
\item we use automatically generated initial points in DAA instead of setting our initial guess. 
\end{itemize} 

\subsection{Values of the Systems Parameters. Statistical Characteristics}

As in \cite{Pechen_Petruhanov_Morzhin_Volkov_CNOT_CPHASE_gates}, consider the Systems~1,~2,~3 with the following values of the systems parameters:
\begin{align*}
\omega_1 = 1, \quad \omega_2 = 1.1, \quad 
\Omega_1 = \Omega_2 = \Lambda_1 = \Lambda_2 = 0.5
\end{align*} 
and the final time $T=20$. For System~3, set the parameter $\alpha = 0.2$. 

Independently for System~1, System~2, and System~3, we consider the series of values 
\begin{align}
\varepsilon \in \{0, ~0.01, ~0.02, ~0.03, ~0.04, ~0.05, ~0.06, ~0.07, ~0.08, ~0.09, ~0.1\}
\label{values_of_epsilon}
\end{align}
and have the problem of minimizing the objective of the general type~(\ref{objective_functional}) for a~given quantum gate for analyzing the corresponding dependences between increasing $\varepsilon$ and the objective values.   

For a~fixed quantum gate~$U$ and the parameter $\varepsilon = \varepsilon_k
\in \{0, 0.01, 0.02, \dots, 0.1\}$, consider the following characteristics over
10 trials of DAA started from automatically generated initial points:
\begin{align}
F^{\rm GRK,sd}_{U,~\text{min-min}}(\varepsilon_k) &= 
\min\limits_{1 \leq i \leq 10} \Big\{ F^{\rm GRK,sd}_U(f_i^{\ast}; \varepsilon_k) \Big\}, \label{J_U_min_min} \\
F^{\rm GRK,sd}_{U,~\text{max-min}}(\varepsilon_k) &= 
\max\limits_{1 \leq i \leq 10} \Big\{ F^{\rm GRK,sd}_U(f_i^{\ast}; \varepsilon_k) \Big\}, \label{J_U_max_min} \\
F^{\rm GRK,sd}_{U,~\text{mean-min}}(\varepsilon_k) &= 
\frac{1}{10} \sum\limits_{i=1}^{10} F^{\rm GRK,sd}_U(f_i^{\ast}; \varepsilon_k), \label{J_U_mean_min}   
\end{align}
where consider $\{f_i^{\ast}\}_{i=1}^{10}$ being the set 
of the resulting piecewise constant controls computed 
via 10~trials of DAA for the same problem of minimizing the objective
$F^{\rm GRK,sd}_U(f)$ under the given~$U,~\varepsilon_k$. Thus, 
the word ``min'' on the right in the lower index reflects this numerical meaning. 

\subsection{For {\rm C-NOT} Gate}

\begin{figure}[ht!]
\centering
\includegraphics[width=0.8\linewidth]{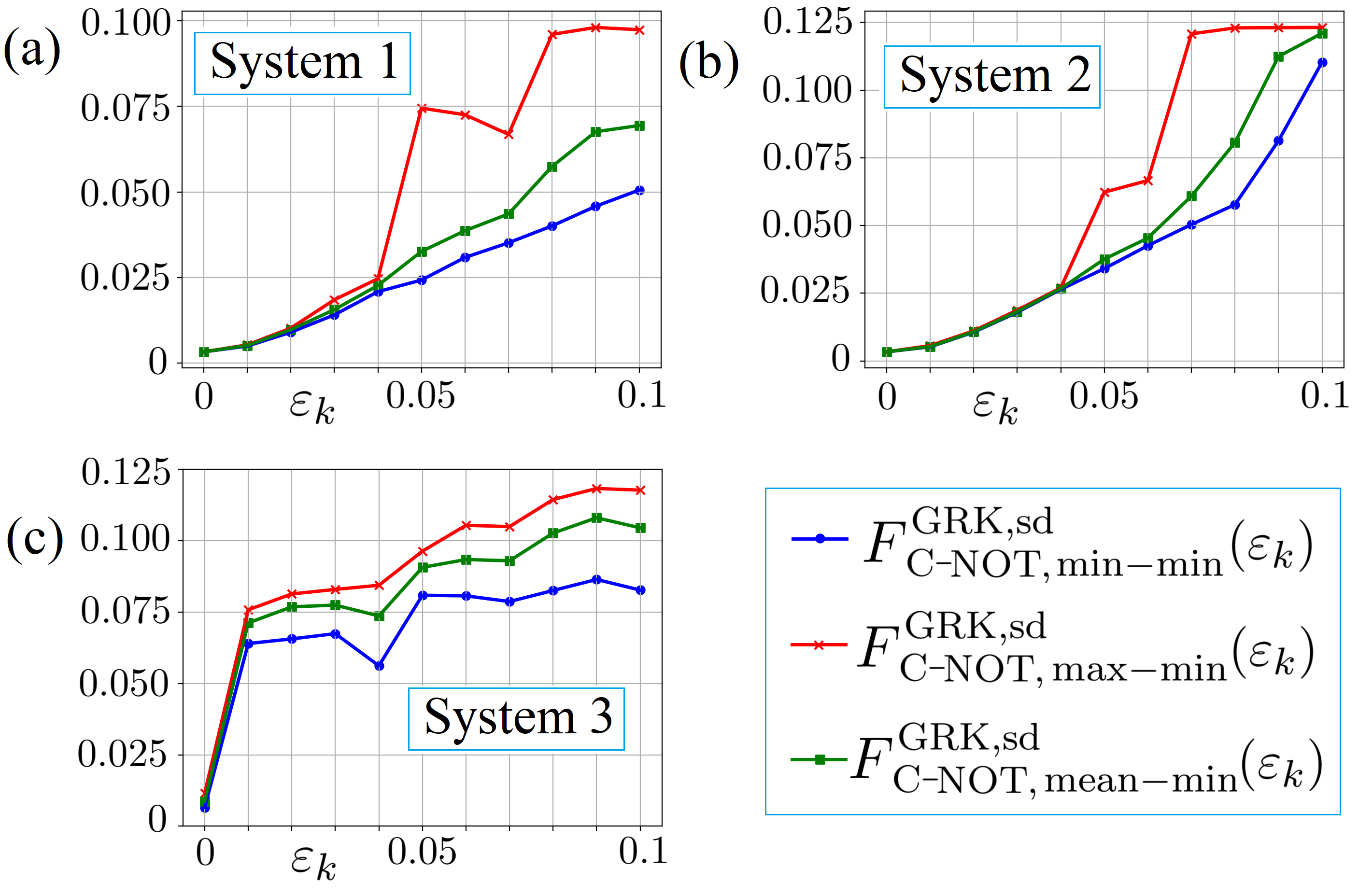} 
\caption{For C-NOT gate and Systems~1,~2,~3, the obtained via DAA values of (\ref{J_U_min_min})--(\ref{J_U_mean_min}): (a)~for System~1; (b)~for System~2; (c)~for System~3. The DAA computed values are shown via the markers, the piecewise linear interpolation is shown in addition. \label{Figure1}} 
\end{figure}
For Systems~1,~2,~3, the results of DAA are shown in~Figure~\ref{Figure1} in terms of (\ref{J_U_min_min})--(\ref{J_U_mean_min}). We see that increasing $\varepsilon_k$ from~0 to~0.1 with the step~0.01  gives, in principle: 1)~increasing the min--min (graphs with round markers), max--min (graphs with square markers), and mean--min  (graphs with x-form markers) values of the objective; 2)~increasing the diversity between the min--min and mean--min, max--min values  for the same~$\varepsilon$, i.e. we see that increasing $\varepsilon$ makes the DAA work more  difficult.  

\begin{figure}[ht!]
\centering
\includegraphics[width=0.8\linewidth]{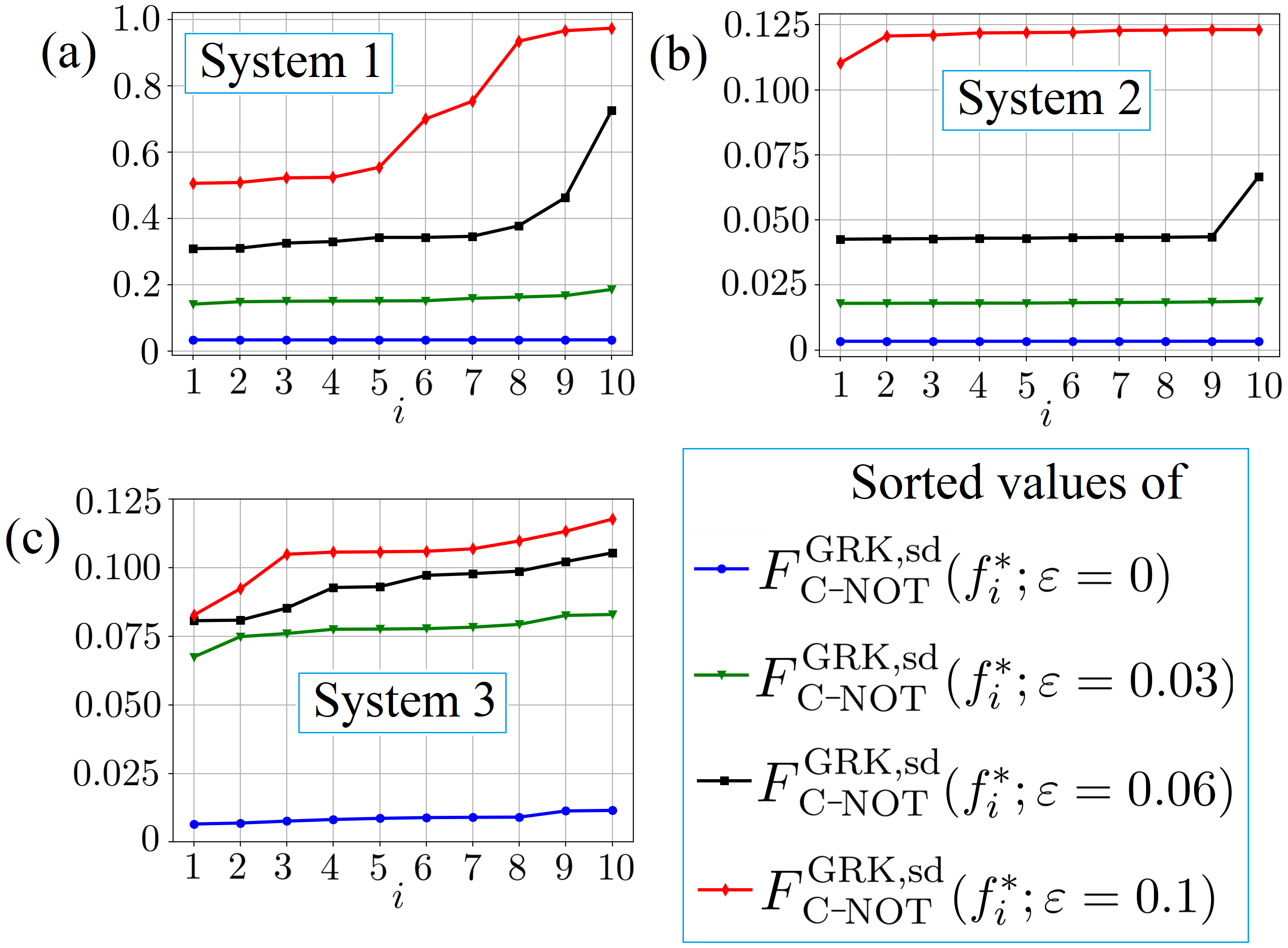} 
\caption{For C-NOT gate and Systems~1,~2,~3. For each $\varepsilon = 0,~0.03,~0.06,~0.1$, the resulting values of the objective $F_{\rm C-NOT}^{\rm FRK,sd}(f)$ in all the 10~trials of DAA: (a)~for System~1; (b)~for System~2; (c)~for System~3. The DAA computed values are shown via the markers. \label{Figure2}} 
\end{figure}

Figure~\ref{Figure2} shows, for each $\varepsilon = 0,~0.03,~0.06,~0.1$, the resulting values of the objective $F_{\rm C-NOT}^{\rm FRK,sd}(f)$ in all the 10~trials of DAA for Systems~1,~2,~3; these objective's values show that: 1)~increasing $\varepsilon$ removes the objective's values further from zero; 2)~DAA used for the same case can give different resulting values of the objective under the same settings of DAA that reminds about the importance of performing at least more than one of trials of DAA with the further comparing. 

For each of System~1, System~2, and System~3, we performed 10 DAA trials for each of 11 values of $\varepsilon$. In each of these 330 DAA trials, we observe which minimal and maximal values of coherent and two incoherent controls considered as piecewise constrant with $K=200$ partitions on~$[0,T=20]$. With respect to the question about incoherent controls, shortly we note that their resulting values in this series of trials are various in the admissible range~$[0, n_{\max}=20]$. 

\subsection{For {\rm SWAP} Gate}

\begin{figure}[ht!]
\centering
\includegraphics[width=0.8\linewidth]{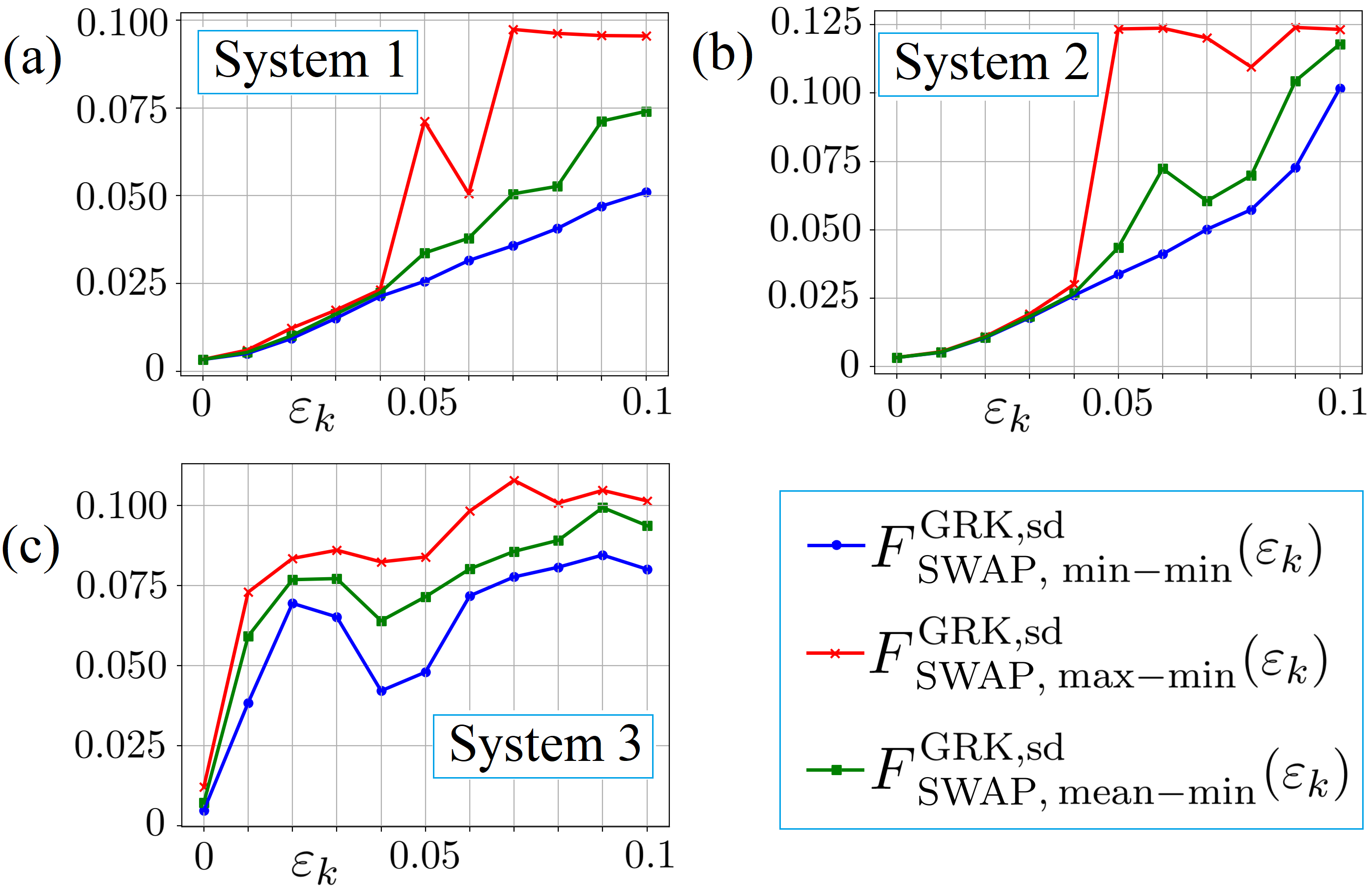} 
\caption{For SWAP gate and Systems~1,~2,~3, the obtained via DAA values of (\ref{J_U_min_min})--(\ref{J_U_mean_min}): (a)~for System~1; (b)~for System~2; (c)~for System~3. The DAA computed values are shown via the markers. \label{Figure3}} 
\end{figure}
\begin{figure}[ht!]
\centering
\includegraphics[width=0.8\linewidth]{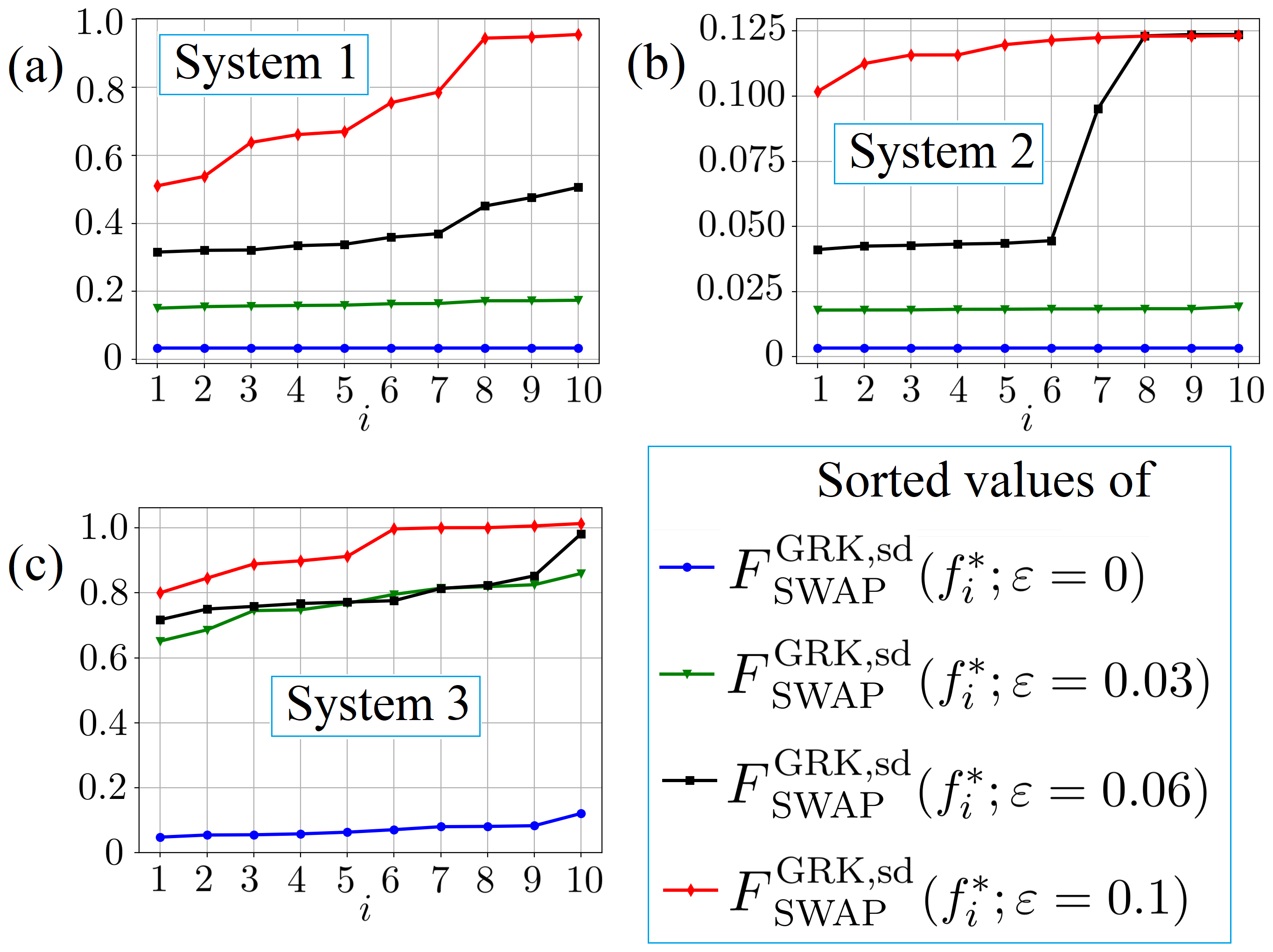}  
\caption{For SWAP gate and Systems~1,~2,~3. For each $\varepsilon = 0,~0.03,~0.06,~0.1$, the resulting values of the objective $F_{\rm C-NOT}^{\rm FRK,sd}(f)$ in all the 10~trials of DAA: (a)~for System~1; (b)~for System~2; (c)~for System~3. The DAA computed values are shown via the markers. \label{Figure4}} 
\end{figure}

For Systems~1,~2,~3, the results of DAA are shown in~Figure~\ref{Figure3} in terms of (\ref{J_U_min_min})--(\ref{J_U_mean_min}). Figure~\ref{Figure4} shows, for each $\varepsilon = 0,~0.03,~0.06,~0.1$, the resulting values of $F_{\rm C-NOT}^{\rm FRK,sd}(f)$ in all the 10~trials of DAA for Systems~1,~2,~3. Figures~\ref{Figure3},~\ref{Figure4} for SWAP gate provide the similar conclusions to what is written above about the dependences with respect to~Figures~\ref{Figure1},~\ref{Figure2} for C-NOT gate. 

\subsection{For {\rm C-Z} Gate}

\begin{figure}[ht!]
\centering
\includegraphics[width=0.8\linewidth]{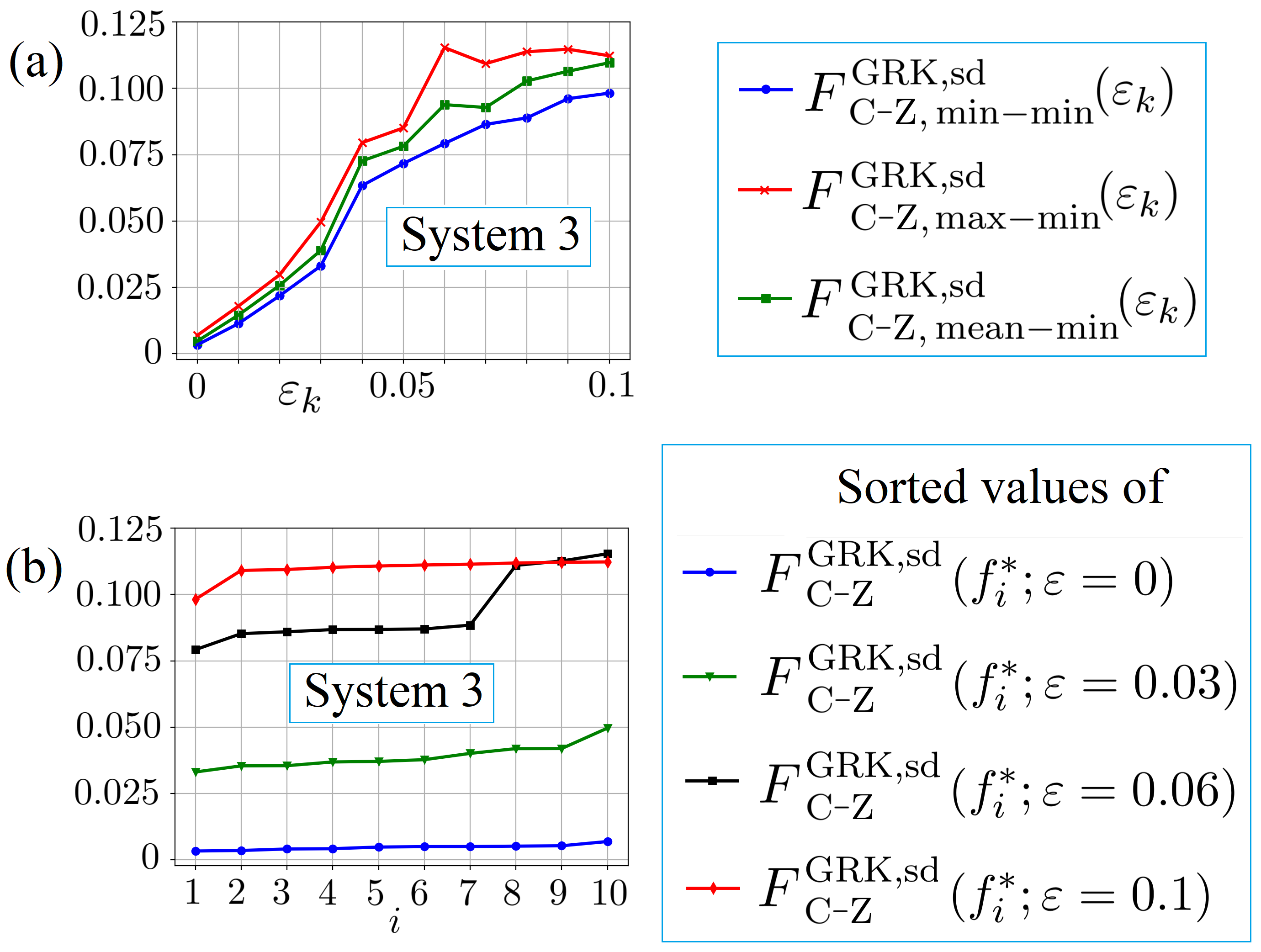}  
\caption{For C-Z gate (i.e. $\pi$-C-PHASE gate) and System~3, the obtained via DAA results: (a)~values of (\ref{J_U_min_min})--(\ref{J_U_mean_min}); (b)~for each $\varepsilon = 0,~0.03,~0.06,~0.1$, the resulting values of the objective $F_{\rm C-NOT}^{\rm FRK,sd}(f)$ in all the 10~trials of DAA. \label{Figure5}} 
\end{figure}

For System~3, the results of DAA are shown in~Figure~\ref{Figure5}(a) in the terms of (\ref{J_U_min_min})--(\ref{J_U_mean_min}) and in~Figure~\ref{Figure5}(b) in the terms of the series of 10~trials of DAA. These figures for C-Z gate provide the similar conclusions to what is written above about the dependences with respect to~Figures~\ref{Figure1}--\ref{Figure4} for C-NOT and SWAP gates. 

\section{Conclusions}
\label{Section_Conclusion}

This work considers a general form of open quantum systems 
determined by the GKSL type 
master equation with simultaneous coherent and incoherent controls 
and the three versions of this general form with three Hamiltonians. Coherent control enters   
through the Hamiltonian and incoherent control enters 
through both the Hamiltonian 
and the superoperator of dissipation. For these systems, 
we analyze the control problems of generating 
two-qubit C-NOT, SWAP, and C-Z gates using piecewise constant controls and stochastic optimization 
in the form of adapted version of the dual annealing algorithm. 
In the numerical experiment, we analyze the dependencies 
between the obtained by the dual annealing minimal objective 
values for various values of the strentgh of interaction between the system and the environment.

The obtained numerical results for 
increasing $\varepsilon$ from~0 to~0.1 with 
the step~0.01 show that: 
\begin{itemize}
\item increasing infidelity further from zero, 
increasing the min--min~(\ref{J_U_min_min}), 
max--min~(\ref{J_U_max_min}), and mean--min ~(\ref{J_U_mean_min}) values of the infidelity;
\item increasing the diversity between the min--min 
and mean--min, max--min infidelity values for the same~$\varepsilon$, 
i.e. we see that increasing $\varepsilon$ makes 
the DAA operation more difficult;
\item DAA operation with the same 
settings and for the same situation can give different resulting values of the objective that reminds about the importance of performing multiple trials  of DAA for a subsequent comparison.
\end{itemize} 

\section*{Funding} 

This work was funded by Russian Federation represented by the Ministry 
of Science and Higher Education of Russian Federation (Project No. 075-15-2020-788).

\section*{Acknowledgments} 

We thank Vadim~N. Petruhanov for his Python code for computing values of 
the objective by operating with the  matrix exponentials corresponding to the realificated quantum systems with piecewise constant controls. This Python code was included by the first author into the code realizing the stochastic optimization approach for the article~\cite{Pechen_Petruhanov_Morzhin_Volkov_CNOT_CPHASE_gates} and this article.

\end{document}